# Moiré patterns of space-filling curves

Henning U. Voss[1,2,*] & Douglas J. Ballon[2]

[1]Cornell MRI Facility, Cornell University, Ithaca, New York, NY, USA
[2]Department of Radiology, Weill Cornell Medicine, New York, NY, USA
*Corresponding author

*Abstract* — It is shown on the examples of Moore and Gosper curves that two spatially shifted or twisted, pre-asymptotic space-filling curves can produce large-scale superstructures akin to moiré patterns. To study physical phenomena emerging from these patterns, a geometrical coupling coefficient based on the Neumann integral is introduced. It is found that moiré patterns appear most defined at the peaks of those coefficients. A physical interpretation of these coefficients as a measure for inductive coupling between radiofrequency resonators leads to a design principle for strongly overlapping resonators with vanishing mutual inductance, which might be interesting for applications in MRI. These findings are demonstrated in graphical, numerical, and physical experiments.

I. INTRODUCTION

Moiré patterns are interference patterns generated by the overlay of a ruled pattern with transparent gaps on another similar pattern. The spatial scale of a moiré pattern is generally larger than that of the individual patterns. Two-dimensional space-filling curves were first discovered in 1890 by Peano [1] and are typically ruled patterns. For example, the plane-filling Hilbert curve can be described by very simple iterative rules [2]. Hilbert pointed out that space-filling curves also define a direction of "motion", forward and backward, in each point of the plane. If one thinks of this motion as a current that flows either way, and connects the two endpoints of such a curve, this defines a magnetic flux through the area enclosed by the curve. In this paper we investigate if pairs of overlaid pre-asymptotic plane-filling curves exhibit a moiré effect, and what influence the emerging large-scale structures have on the flux coupling between the curves.

The original interest in this work arose from the recently emerging importance of moiré patterns in the electrical properties of twisted graphene bilayers [3] and potential translation into the classical domain. Whereas the present scenario is different, as the moiré patterns emerge from a shift or twist between aperiodic structures with only approximative translational or rotational symmetry, we suggest that the present study can provide insights for the design of interesting properties of flux-coupled structures, for example in applications such as magnetic resonance imaging (MRI). MRI sensors are quite generally made of electromagnetic resonator arrays confined to a surface, and the understanding of electromagnetic coupling between array elements plays a crucial role in their engineering [4-8]. Notably, the recently discovered concept of topological modes can arise both in coupled resonator systems [9] as well as in twisted bilayers [10].

The organization of this article is as follows: First, the moiré effect is demonstrated in two different space-filling curves: Spatially shifted Moore curves, and twisted Gosper curves. These two examples represent curves with approximate translational and rotational symmetry, respectively. Next, the Neumann integral is introduced as a geometric measure of distance in space-filling curves, and it is shown that its value peaks at those shifts or twists with the most defined moiré effect. We then demonstrate in a physical experiment that the interpretation of the Neumann integral as a measure for inductive coupling in radiofrequency resonators leads to design rules for strongly overlapping space-filling resonators with vanishing mutual inductance. Finally, limitations of the Neumann integral formulation and potential applications are discussed.

## A. Graphical experiments: Moiré patterns of the Moore and Gosper curves

In the first graphical experiment, we investigate the large-scale structures emerging from pre-asymptotic Moore curves (Figure 1A). The Moore curve is a generalization of the Hilbert curve. In the Hilbert curve, the start and end points are at different corners, and the Moore curve brings them next to each other by stitching four Hilbert curves together. Connecting start and end points then forms a closed loop, such that current passing through it will induce a magnetic flux.

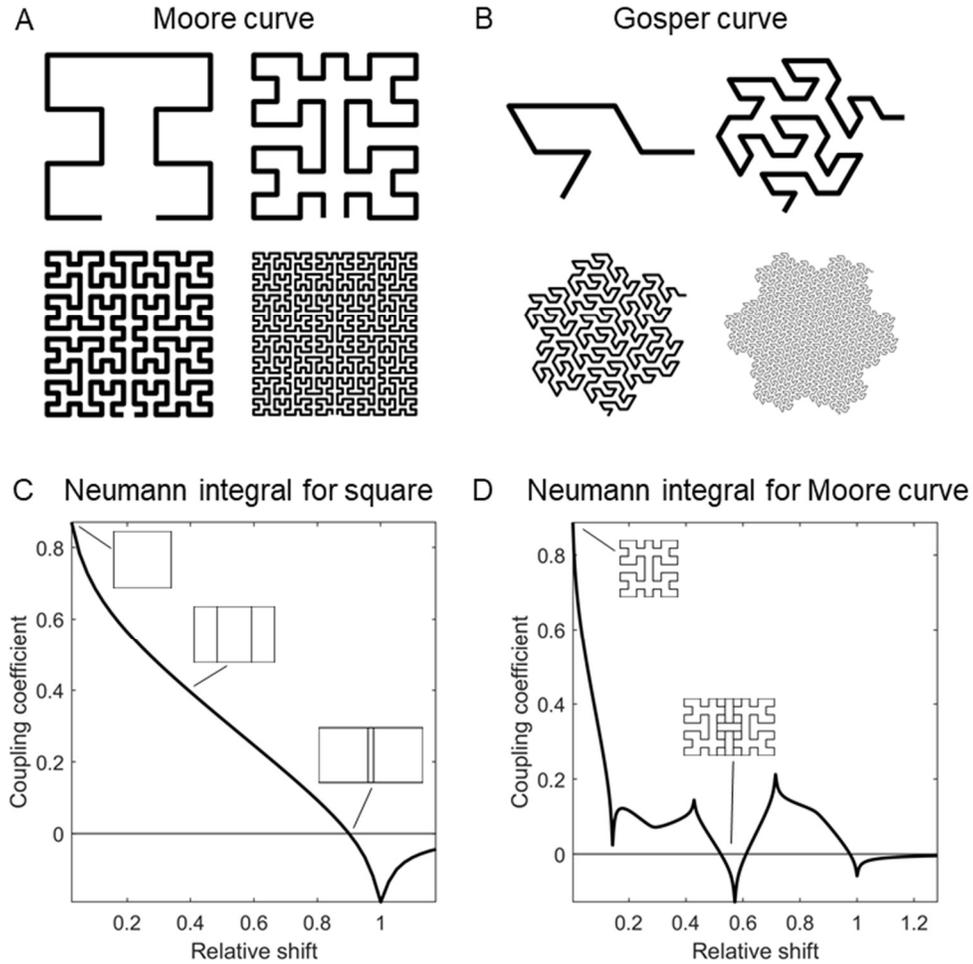

**Figure 1: Space filling curves and Neumann integrals.** A) The first four iterations of the Moore curve. B) The first four iterations of the Gosper "snowflake" curve. C) The Neumann integral for shifted squares ($\Delta = 0.001$). Square resonators have vanishing inductive coupling at a relative shift of about 0.92. D) The Neumann integral for shifted Moore curves with $N_i = 2$ ($\Delta = 0.01$). Moore resonators with $N_i = 2$ have vanishing inductive coupling at relative shifts of 0.52, 0.61, and 0.97.

Moore curves were generated with MATLAB (The Mathworks, version R2023a), using a Lindenmayer algorithm [11, 12]. The number of iterations of the Lindenmayer system was $N_i = 7$. Figure 2 shows examples of the moiré effect in these curves. For certain shifts, superstructures emerge, consisting of parts that locally resemble the original Moore curves and parts that appear more complex.

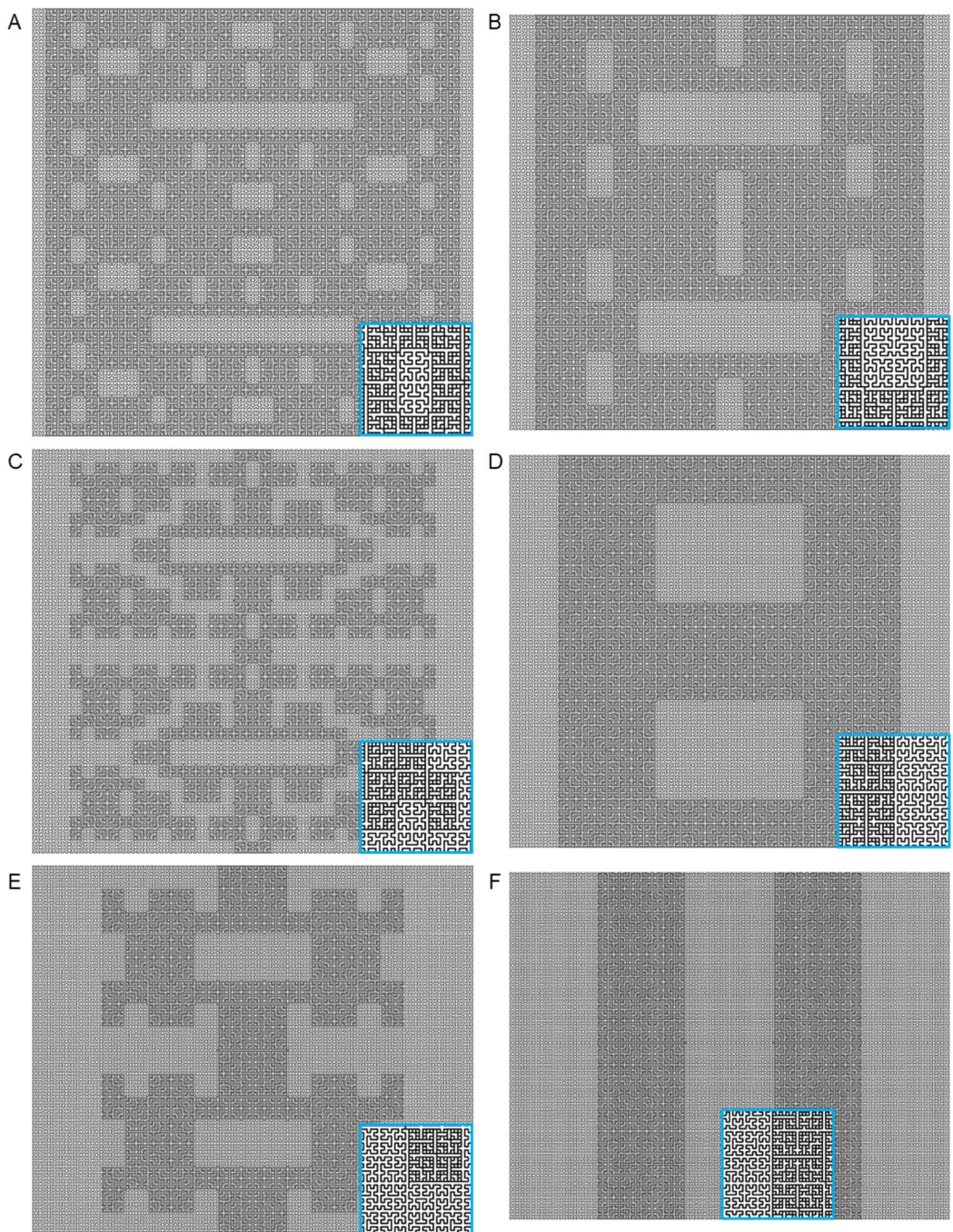

**Figure 2: Moiré effect in space-filling Moore curves.** Each curve is shown with a magnified inset to reveal its details. The relative spatial shifts of the two Moore curves are A) 3.1%, B) 6.3%, C) 9.4%, D) 13%, E) 19%, F) 25%.

The second example demonstrates the moiré effect in space-filling curves exhibiting approximate rotational rather than translational symmetry. We chose the Gosper "snowflake" curve (Figure 1B) [13], which has 30° and 60° intrinsic angles that follow from the curve's construction by connecting nodes of a hexagonal lattice. For this case of approximate rotational symmetry, it is natural to anticipate twisted moiré patterns. Figure 3 shows examples for the moiré effect in these curves with $N_i = 6$ for selected twist angles. Indeed, for certain angles, superstructures with six-fold symmetry emerge. Again, the details consist of parts that locally resemble the original Gosper curves and parts that appear more complex.

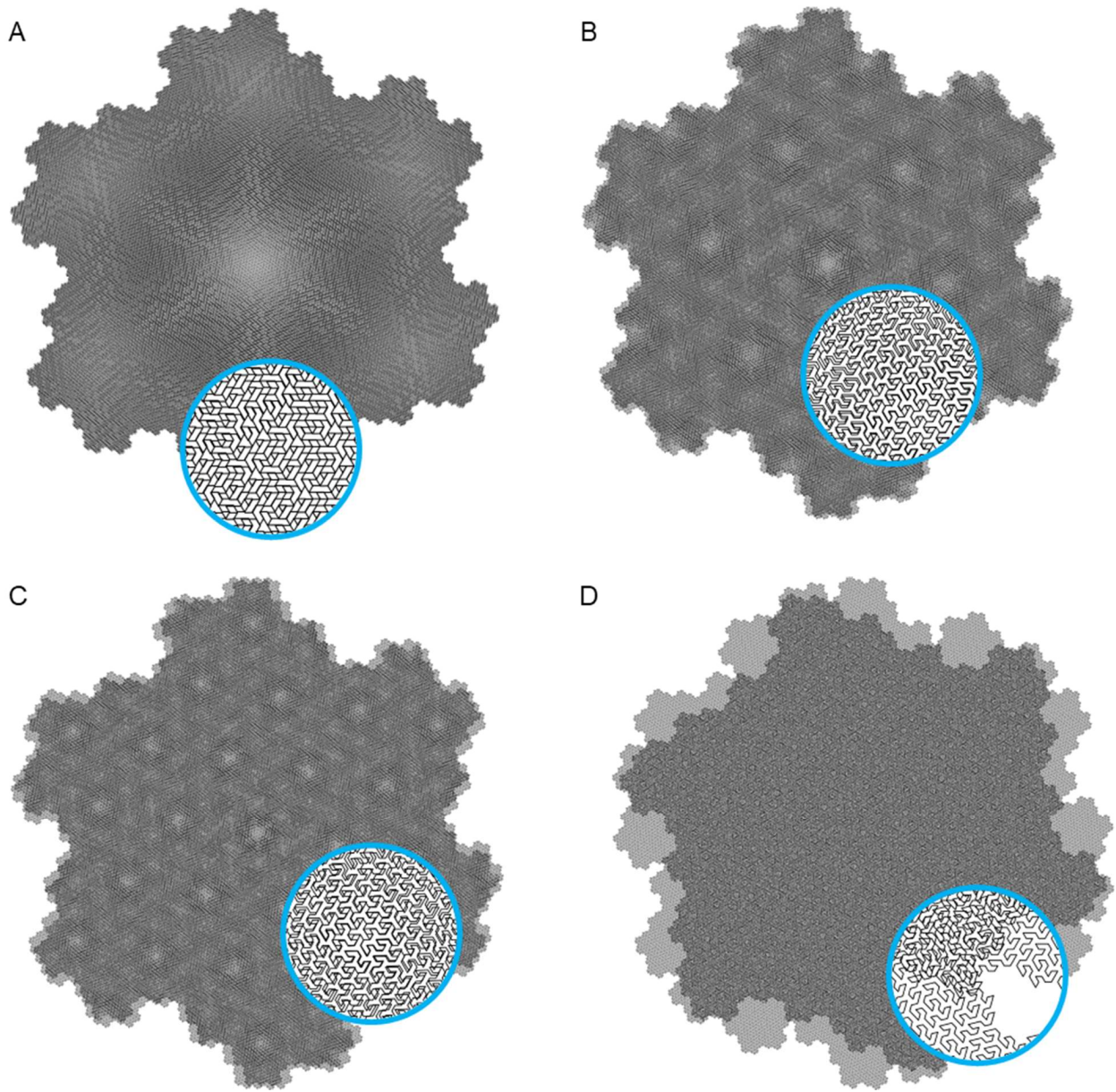

**Figure 3: Moiré effect in space-filling Gosper curves.** Each curve is shown with a magnified inset to reveal its details. The rotation angles between the two Gosper curves are A) 0.4º, B) 1.6º, C) 2.4º, D) 14.2º.

## B. Numerical experiments: Coupling analysis of moiré patterns

The Neumann integral [14] is a geometric measure [15] of distance between curves with the property that the directionality of the curve is taken into account. This property corresponds to the aforementioned "motion" directionality in space-filling curves by Hilbert. With proper physical scaling, the Neumann integral approximates the mutual inductance between two closed loops made of thin wires with radius negligible compared to their lengths. The normalized Neumann integral provides a direct estimate of the inductive coupling coefficient, a dimensionless quantity that will be used here.

The normalized Neumann integral for our application of two spatially shifted in-plane curves is defined as

$$\zeta_\alpha = \frac{1}{N_0} \oint_{C_1} \oint_{C_{2(\alpha)}} \frac{d\boldsymbol{r}_1 \cdot d\boldsymbol{r}_2}{|\boldsymbol{r}_1 - \boldsymbol{r}_2|} \; , \tag{1}$$

where $C_1$ and $C_2$ are the curves defined by the conductors, α is the shift distance or twist angle between the two curves, and *dr₁* and *dr₂* are the infinitesimal increments of vectors *r₁* and *r₂* defining the curves. It is assumed that the two curves have a small, constant offset Δ perpendicular to the plane to avoid divergence for α = 0 and to provide a realistic approximation for conductors on a substrate with finite thickness. The normalization constant $N_0$ is obtained by the same double line integral for α = 0.

The notion of the Neuman integral as a geometric distance measure is illustrated here on two identical square loops in the plane, of which one of them has a small spatial offset Δ. Neumann integrals were computed with MATLAB by parallelized summation of line elements. If the square loops are overlapping completely (α = 0), the normalized Neumann integral $\zeta_\alpha$ is one. It decreases for larger α and vanishes once it reaches a point of little overlap at a relative shift of α = 0.92, before it becomes negative (Figure 1C). For larger α, it asymptotically vanishes again. The physical interpretation of this graph is as follows. If the two squares define wires with an inserted capacitor, they model two individual LC resonators that are inductively coupled to each other. Generally, if the two resonators are facing each other in a solenoidal configuration, their coupling (or mutual inductance entering the Kirchhoff equations, including its sign) is positive. If they are oriented in the same plane as in the figure without overlap, their coupling is negative. Upon bringing the loops closer together their coupling first increases in magnitude but finally decreases until it vanishes at α = 0.92 [4]. To understand this behavior, one can simply consider magnetic flux lines emanating from the left loop upwards; if there is a large overlap of the two loops, most flux lines enter the right loop in upward direction, too. For larger shift, or decreasing overlap, the flux lines from the left loop ultimately reverse sign, now entering the right loop from above. These two cases are modeled by a positive and negative coupling coefficient, respectively. The Neumann integrals for two shifted Moore curves are markedly different from the case of two shifted squares (Figure 1D). This will be explored in the following.

Figure 4 shows the computed coupling coefficient of two spatially shifted Moore curves ($N_i$ = 5, Δ = 0.01) as a function of the relative shift α. The Neumann integral or coupling coefficients attain both positive and negative values, and their graph has pronounced peaks. The insets depict selected moiré patterns of the two overlaid Moore curves. Most pronounced moiré effects occur at the peaks of the coupling.

Figure 5 shows the computed coupling coefficient of two twisted Gosper curves ($N_i$ = 4, Δ = 0.001) as a function of the twist angle α, with some selected moiré patterns. The coupling is symmetric around α = 180°, and values for α > 190° are not shown. The rotation center was chosen such that for α = 120° maximum congruence of the two curves was achieved. Overall, the coupling coefficients change in a more complex way with α than in the Moore curve. It was also observed that the curve depends more sensitively on the value of the distance Δ.

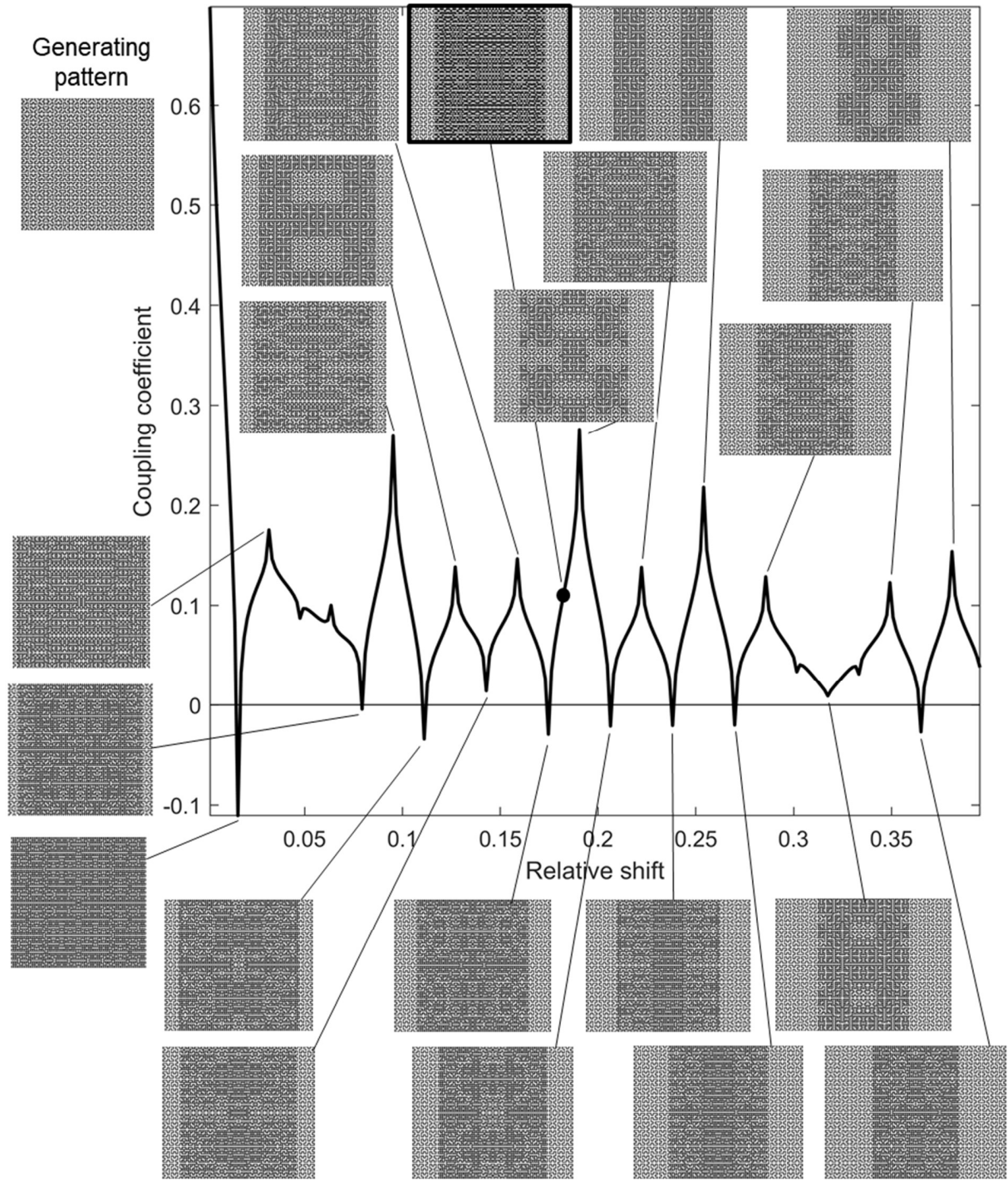

**Figure 4. Coupling coefficients of two shifted Moore curves** for the fifth iteration towards the space-filing curve, as a function of the relative spatial shift. In addition, some of the resulting patterns are shown. The moiré effect appears most defined at the peaks of the coupling coefficients. For the framed pattern, located in-between two peaks, large scale structures are not clearly as discernible anymore.

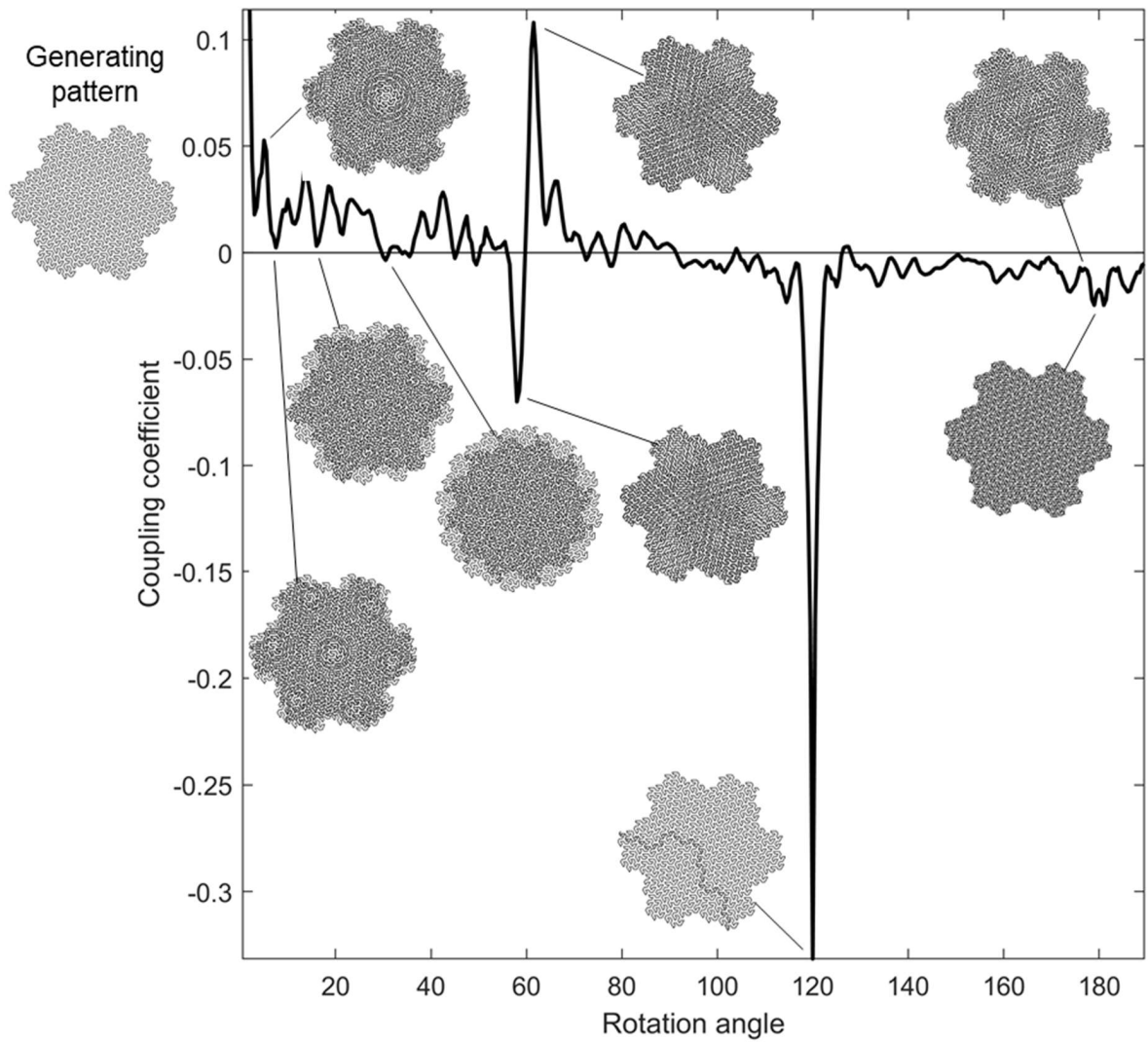

**Figure 5. Coupling coefficients of two twisted Gosper curves** for the fourth iteration towards the space-filing curve, as a function of the rotation angle from 5° to 190°. In addition, some resulting patterns, including moiré patterns, are shown.

## C. Physical experiment: Coupled resonators

For experimental testing of the preceding model simulations, the Moore curve was chosen due to its simpler coupling coefficient behavior under changing spatial shift compared to the more complex Gosper curve's behavior under changing twist angle.

Two Moore radiofrequency resonators with $N_i = 3$ and dimension 7.5 × 7.5 cm were fabricated on a printed circuit board with 1.7 mm thickness and an isolation layer above the copper traces (Figure 6A). A 10 pF-capacitor (measured value 11.7 pF) was inserted serially into one of the corners of the first Moore curve, and a variable capacitor of 20 pF into the second Moore curve. On the first circuit, the measured resistance was 0.63 Ω, and the resonance frequency 55.0 MHz. The second circuit's resonance frequency was matched to the one of the first circuit. The two circuit boards were then aligned while facing each other ($\Delta = 0.2$ mm).

The $S_{11}$ reflection coefficient spectrum of the coupled Moore resonators was measured with a network analyzer (NanoVNA H2/H4, Taobao, Hangzhou, China) for shifts from 17% to 110%, using an inductively coupled pickup loop of about the same size as the resonators. Mode splitting of two coupled LC radiofrequency resonators is expected to be $\omega_{1,2} = \sqrt{(1 \mp \kappa)LC}^{-1}$, from which the coupling coefficient follows as $\kappa = \frac{\omega_1^2 - \omega_2^2}{\omega_1^2 + \omega_2^2}$. Due to resistance, modes have a finite linewidth, here found to be Q = ω/Δω = 55, and zero-coupling shifts were defined as those for which a visible split of resonances could not be observed.

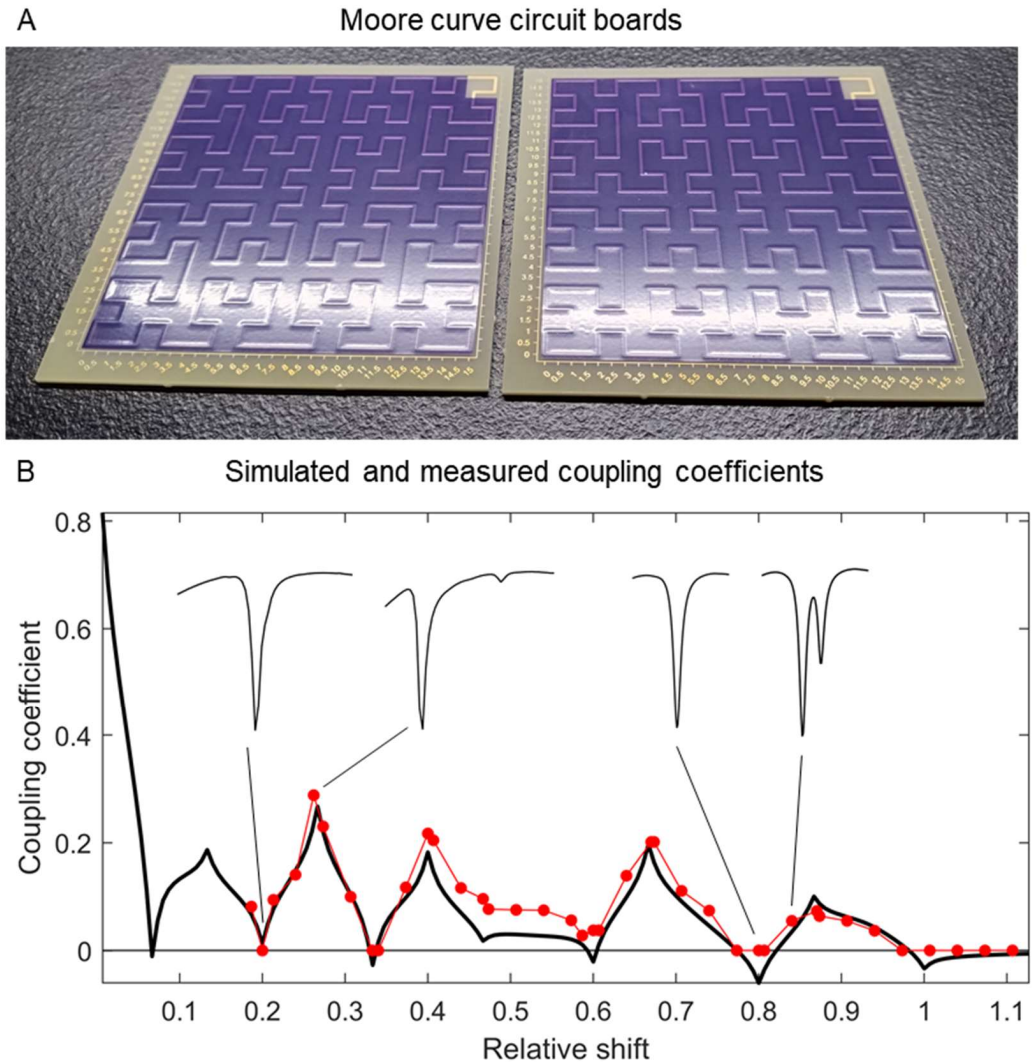

**Figure 6. Physical experiment of coupling in Moore resonators.** A) The copper layers on two printed circuit boards define two Moore curves with $N_i = 3$. After the addition of a capacitor (not shown) they become radiofrequency resonators. B) Observed coupling coefficients (red) vs. Neumann integral simulation (black) for shifted circuit boards. The insets provide examples for single resonances, or zero coupling (first and third inset), and resonance splitting due to inductive coupling (second and fourth inset).

The numerically estimated Moore curve coupling coefficients are shown in Figure 6B, black graph. Experimentally, the shift between the two boards was gradually increased, and the resonance frequencies were obtained from the $S_{11}$ spectrum. From the resonance frequencies, the coupling coefficients κ were computed (Figure 6B, red graph). When the two resonances could not be isolated, the κ = 0 condition is satisfied. This is demonstrated in the first and third insets, which show $S_{11}$ spectra with only one resonance. The second and fourth insets provide examples for $S_{11}$ spectra with split resonances due to nonzero coupling.

Overall, the measurements are in good agreement with the numerical values. Experimental factors such as finite copper trace widths, finite line widths due to resistance, and finite measurement resolution of the network analyzer might contribute to remaining deviations from theory. The near-zero coupling point at relative shifts of about 0.47 could not be isolated with this experiment. In summary, approximate zero coupling could be achieved for significantly overlapping Moore resonators at multiple shift values.

## D. Discussion

Space-filling curves are not only of mathematical interest but have found application in various fields of science and engineering. For example, they have been utilized for decades to study the folding principles of proteins and polymers [16]. Hilbert and Moore curves have been suggested as solutions to build miniature antennas by compressing long wires into small areas [17, 18]. Our interest is in potential applications as radiofrequency resonators. Radiofrequency resonator arrays are the main component of MRI phased arrays [4], for which one usually tries to minimize coupling. In phased arrays, it is important to understand the effects of coupling on the resonance spectra and eigenmodes of the arrays. Sometimes, inductive coupling is even desired in MRI coils such as high-pass surface resonator arrays [6]. The Neumann integral approximates the mutual inductance between closed wire loops under certain conditions, such as thin wires, smoothness, no concave areas, and low frequencies [15]. These conditions are only partially fulfilled in space-filling curve geometries [19, 20] and MRI applications, and would introduce systematic errors in estimates of mutual inductance. Nevertheless, this does not affect the use of the Neumann integral as a geometric distance measure; as we have shown, the Neumann integral peaks at the emergence of moiré patterns, and if used as an estimate of mutual inductance, can indeed predict vanishing inductive coupling configurations of space-filling radiofrequency resonators.

## E. Summary

It has been demonstrated that pre-asymptotic space-filling curves naturally lead to moiré patterns. Moiré patterns emerge in shifted space-filling curves with approximate translational symmetry as well as in twisted space-filling curves with approximate rotational symmetry. The Neumann integral was used as a geometric distance measure between two space-filling curves, and it was shown that it peaks at those shifts or twists that cause a pronounced moiré effect. Furthermore, if interpreted as a measure of inductive coupling, the Neumann integral also predicts zero-coupling configurations of space-filling radiofrequency resonators, an important design criterion for potential MRI applications.

## F. Disclosure

Disclosure: The authors are inventors on patents, owned by Cornell University, that are related to radiofrequency resonators akin to those in this manuscript.

___